\newif\ifarxiv 
\arxivtrue 

\ifarxiv\documentclass[aps,pra,twocolumn,superscriptaddress,showpacs,nofootinbib]{revtex4-2}\else\documentclass[aps,pra,twocolumn,superscriptaddress,showpacs]{revtex4-2}\fi

\usepackage{amsmath,amssymb,amsfonts}
\usepackage{graphicx, subcaption}
\usepackage{bm}
\usepackage{braket}
\usepackage{hyperref}
\usepackage{cancel}
\usepackage{booktabs}
\usepackage{multirow}
\usepackage[usenames,dvipsnames]{xcolor}
\usepackage[normalem]{ulem}

\begin{document}

\title{Randomness can be certified in energy-constrained semi-device-independent scenarios}

\author{Shashank Kumar Ranu}
\email{shashank-kumar.ranu@inria.fr}
\affiliation{Universit\'e Grenoble Alpes, Inria, 38000 Grenoble, France}
\author{Lewis Wooltorton}
\email{lewis.wooltorton@ens-lyon.fr}
\affiliation{Inria, ENS de Lyon, UCBL, LIP, 46 Allee d’Italie, 69364 Lyon Cedex 07, France}
\author{Alastair A.\ Abbott}
\email{alastair.abbott@inria.fr}
\affiliation{Universit\'e Grenoble Alpes, Inria, 38000 Grenoble, France}
\author{Omar Fawzi}
\email{omar.fawzi@ens-lyon.fr}
\affiliation{Inria, ENS de Lyon, UCBL, LIP, 46 Allee d’Italie, 69364 Lyon Cedex 07, France}
\date{\today}

\begin{abstract}
The prepare-and-measure framework based on energy constraints offers a practical middle ground between fully device-dependent and device-independent quantum cryptography. The only assumption on an otherwise uncharacterized prepare-and-measure device is that the energy of the prepared states is bounded. Existing security analyses of this framework assume that the preparation and measurement devices share at most classical correlations, and under this assumption certified lower bounds on the extractable randomness have been established. Recent work has shown that an adversary who pre-distributes entanglement between the devices can mount attacks that are strictly more powerful than those available when the devices share only classical correlations, reducing the extractable randomness below the previously certified rates. This leaves open the fundamental question of whether randomness can be certified at all in this scenario. 

We address this open question by constructing semidefinite programming relaxations of the guessing probability by adapting the Navascués-Pironio-Acín hierarchy to the energy-constrained setting where shared entanglement between the devices is permitted. Our relaxations yield certified lower bounds on the extractable randomness without enforcing any restrictions on the dimensions of the quantum state shared between the preparation and measurement devices. We show that these certified lower bounds are strictly positive for a range of energy values, thereby answering the open question affirmatively: certified randomness generation is theoretically possible in the energy-constrained SDI framework even in the presence of a fully quantum adversary.
\end{abstract}

\maketitle

\ifarxiv\section{Introduction}\else\textit{Introduction.---}\fi Quantum physics enables the generation of intrinsically random numbers whose unpredictability is guaranteed by fundamental physical laws. The security of such quantum random number generators (QRNGs) is analyzed under different levels of trust in the devices. At one end of the spectrum, device-dependent QRNGs establish security under the assumption that the state preparation and measurement devices are completely characterized. This device-dependent assumption permits simple and fast implementations~\cite{bruynsteen2023100,li2026high,hua2026fully}. However, ensuring complete knowledge of the devices in practice is difficult, and any deviation of the physical hardware from its theoretical model can compromise the security guarantees. At the other end of the spectrum,  device-independent (DI) QRNGs treat the devices as black boxes, certifying randomness solely from observed input–output statistics~\cite{colbeck2009quantum, pironio2010random}. While DI QRNGs offer the strongest form of security, it comes at the price of stringent experimental requirements. DI QRNG implementations must simultaneously close the detection and locality loopholes, which severely limits achievable generation rates with current technology~\cite{liu2018device,liu2021device,bierhorst2018experimentally,shalm2021device}.

The semi-device-independent (SDI) framework offers a practical middle ground between these two extremes. It retains the prepare-and-measure (PM) architecture of device-dependent protocols, avoiding the need for entanglement distribution and loophole-free Bell tests, while requiring only a minimal set of assumptions on the otherwise uncharacterized devices~\cite{brask2026quantum}. Several such assumptions have been explored, including bounds on the Hilbert-space dimension of the prepared states~\cite{li2011semi, lunghi2015self, li2012semi}, constraints on their pairwise overlaps~\cite{brask2017megahertz}, fidelity bounds with respect to target states~\cite{tavakoli2021semi}, and energy constraints~\cite{van2017semi,van2019correlations}. Among these, the energy constraint is particularly appealing from an experimental standpoint, as in photonic implementations it corresponds to a bound on the mean photon number, a quantity that can be monitored using standard off-the-shelf optical components. This practical advantage has led to several fast SDI QRNG implementations based on energy constraints~\cite{avesani2021semi, rusca2020fast,sabatini2026efficient}. 

However, all existing energy-constrained SDI QRNG implementations have been analyzed under a restrictive eavesdropping model, in which the preparation device (Alice) and the measurement device (Bob) share at most classical correlations, typically described by a random variable $\lambda$ known to the adversary~\cite{van2017semi,van2019correlations,rusca2020fast,sabatini2026efficient}. Under this assumption, certified lower bounds on the extractable randomness have been obtained in both the asymptotic and finite-round regimes~\cite{van2019correlations,sabatini2026efficient, bhavsar2026higher}. Recent works have investigated the impact of lifting this assumption, considering the more general scenario in which Alice and Bob may share quantum entanglement, with the adversary holding a purification of their joint state~\cite{d2025entanglement,roch2026role}.  
These works showed that entanglement strictly enlarges the set of input--output statistics achievable under the energy constraint and can substantially reduce the certifiable randomness. This is concluded by heuristically searching for explicit eavesdropping attacks that outperform those available to an adversary restricted to distributing classical correlations between the devices. However, this approach falls short of what is required for a security proof, namely, a lower bound on the generated randomness that is valid for \emph{all} attacks compatible with observations. This leaves open the question of whether randomness can be certified in this scenario.

In this \ifarxiv{paper}\else{letter}\fi, we answer this question affirmatively. We construct semidefinite programming (SDP) relaxations of the guessing probability based on the Navascués-Pironio-Acín (NPA) hierarchy~\cite{NPA1,navascues2008convergent}, adapted to the energy-constrained PM scenario in which the devices may share arbitrary entanglement. Unlike the previous upper bounds, which impose dimension restrictions on the shared entangled state, our approach provides a certified lower bound on the randomness rates that hold in any, arbitrarily large, finite dimension for the underlying systems. Equipped with this, we show that a non-zero rate of randomness generation can be achieved for a range of experimental parameters. Furthermore, we show that if one is additionally willing to assume a bound on the source dimension, our technique can be modified to provide certified rates that nearly coincide with the upper bounds of Refs.~\cite{d2025entanglement,roch2026role}.

\begin{figure*}[t]
    \centering
    \begin{subfigure}[t]{0.4\textwidth}
        \centering
        \includegraphics[width=0.8\linewidth]{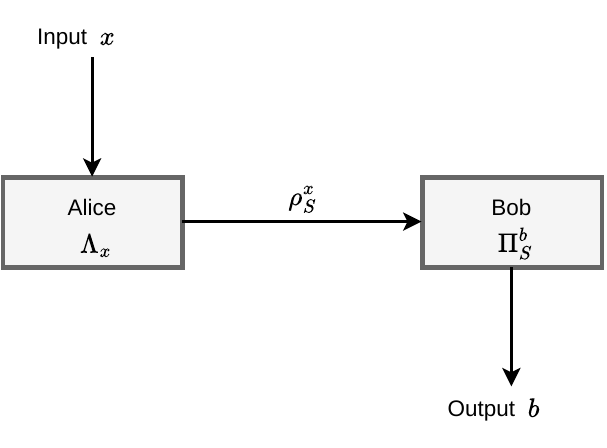}
        \caption{}
        \label{fig:honest}
    \end{subfigure}
    \hspace{0.02\textwidth}
    \begin{subfigure}[t]{0.4\textwidth}
        \centering
        \includegraphics[width=0.8\linewidth]{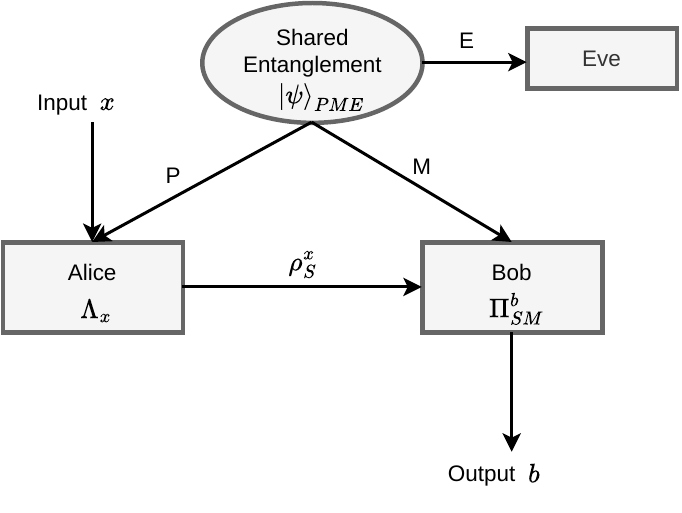}
        \caption{}
        \label{fig:EAPM}
    \end{subfigure}

    \caption{Energy-constrained prepare-and-measure scenario (a) honest implementation and
    (b) adversarial implementation with shared entanglement.}
    \label{fig:PM}
\end{figure*}

\ifarxiv\section{Energy-constrained PM scenario with shared entanglement}\else\textit{Energy-constrained PM scenario with shared entanglement.---}\fi We consider a minimal prepare-and-measure scenario for randomness generation with a fully quantum adversary. In an honest implementation (see Fig.~\ref{fig:honest}), the preparation (Alice) and measurement (Bob) devices are uncorrelated. Alice encodes her input into an energy-constrained quantum state, which she sends to Bob, who measures it. Our analysis, however, covers the more general adversarial scenario (see Fig.~\ref{fig:EAPM}) in which the adversary (Eve) prepares a global pure state $\ket{\psi}_{PME}$, distributing systems $P$ and $M$ to the preparation and measurement devices while retaining the purifying system $E$. Let $\sigma_{PM} := \operatorname{Tr}_E[\ket{\psi}\!\bra{\psi}_{PME}] \in \mathcal{D}(\mathcal{H}_P \otimes \mathcal{H}_M)$ denote this initial shared state\ifarxiv{}\else{~}\fi\footnote{We denote by $\mathcal{D}(\mathcal{H})$ the set of density operators (positive semidefinite with unit trace) on a Hilbert space $\mathcal{H}$, and by $\mathcal{L}(\mathcal{H})$ the set of linear operators on $\mathcal{H}$. The identity channel from $\mathcal{L}(\mathcal{H})$ to itself is denoted by $\mathcal{I}$.} between Alice and Bob, which may in general be entangled. In each round of the protocol, Alice receives a uniformly random input $x \in \{0,1\}$ and applies a quantum channel $\Lambda_x : \mathcal{L}(\mathcal{H}_P) \to \mathcal{L}(\mathcal{H}_S)$ to her share of $\sigma_{PM}$, producing the joint state $\rho^x_{SM} = (\Lambda_x \otimes \mathcal{I}_M)\,[\sigma_{PM}]$, whose marginal on the source system is $\rho^x_S = \operatorname{Tr}_M[\rho^x_{SM}]$. She sends $\rho^x_S$ to Bob, who performs a single binary measurement described by a POVM $\{\Pi^b_{SM}\}_{b \in \{0,1\}}$ on the joint system $SM$ and obtains the outcome $b$. The resulting input--output statistics are characterized by the conditional probabilities $p(b|x) = \operatorname{Tr}[\Pi^b_{SM}\,\rho^x_{SM}]$.

The semi-device-independent character of the above PM scenario comes from a single assumption on the prepared states, i.e., they must have a large overlap with a common ground state.  Formally, there exists a state $\ket{0} \in \mathcal{H}_S$ such that
\begin{equation}\label{eq:energy}
  \bra{0} \rho^x_S \ket{0}
  \;\geq\; 1 - \omega_x
  \qquad \forall\, x \in \{0,1\}
\end{equation}
where $\omega_x \in [0, 1/2)$ is the energy parameter. In photonic implementations, $\ket{0}$ typically is the vacuum, and ensuring that the mean photon number remains below $\omega$ is sufficient to guarantee Eq.~\eqref{eq:energy}. Throughout this work, we assume for simplicity equal energy bounds for both preparations, i.e.,\ $\omega_0 = \omega_1 = \omega$.  

The input--output statistics are equivalently described by the correlators
\begin{equation}\label{eq:correlator_def}
  E_x \;=\; p(0|x) - p(1|x),
  \qquad x \in \{0,1\},
\end{equation}
so that $E_x \in [-1,1]$.  The relevant figure of merit is the correlator difference, $I_{\mathrm{corr}}\;=\;|E_0 - E_1|\,$, which quantifies Bob's ability to distinguish the two prepared states. $I_{\mathrm{corr}}$ can attain its algebraic maximum of $2$ whenever the two preparations are perfectly distinguishable, and it becomes a witness of nonclassicality only in conjunction with the energy assumption of Eq.~\eqref{eq:energy}, which forces both states to have large overlaps with the same ground state and thus limits their distinguishability. Under this constrained energy assumption, the value of $I_{\mathrm{corr}}$ witnesses a hierarchy of resources. For classical strategies\ifarxiv{}\else{~}\fi\footnote{A strategy is classical if it induces a convex combination of deterministic correlations~\cite{van2017semi}, $p(b|x) = \sum_{\lambda}p_{\lambda}\,p(b|x,\lambda)$ where $\lambda$ is a shared random variable, $p_{\lambda}$ its probability distribution and each response function $p(b|x,\lambda) \in \{0,1\}$ is deterministic and realizable by preparations and measurements obeying Eq.~\eqref{eq:energy} with energy parameters $\omega^\lambda_x$, subject to $\sum_\lambda p_\lambda \omega^\lambda_x \leq \omega_x$.}, for which no randomness can be certified, one has $I_{\mathrm{corr}}\leq 4\omega$. When the devices are allowed to share classical randomness, while Alice prepares quantum states and Bob performs a quantum measurement on the received system, $I_{\mathrm{corr}}$ can exceed the classical threshold of $4\omega$, thereby enabling quantum randomness certification. Nevertheless, within this classically correlated model, the correlator is upper-bounded as~\cite{van2017semi,van2019correlations}
\begin{equation}\label{eq:classical_bound}
  I_{\mathrm{corr}}
  \;\leq\; I^{\mathrm{rand}}_{\mathrm{corr}}(\omega)
  \;:=\; 4\sqrt{\omega(1-\omega)}\,.
\end{equation}
A violation $I_{\mathrm{corr}} > I^{\mathrm{rand}}_{\mathrm{corr}}(\omega)$ can be achieved only when entanglement between Alice and Bob is allowed~\cite{d2025entanglement,roch2026role}.

\ifarxiv\section{Randomness generation in the energy-constrained SDI scenario with shared entanglement}\else\textit{Randomness generation in the energy-constrained SDI scenario with shared entanglement.---}\fi We employ a spot-checking protocol for randomness generation in this energy-constrained PM scenario. In the majority of rounds, designated generation rounds, Alice's input is fixed to $x = 0$, preparing the state $\rho_S^0$. Bob measures the corresponding state $\rho^0_{SM}$ using the binary POVM $\{\Pi^b_{SM}\}_{b \in \{0,1\}}$, and the outcome $b$ constitutes a raw random bit. In the remaining small fraction of rounds, designated test rounds, Alice chooses her input uniformly at random from $\{0,1\}$. Bob collects the input--output statistics $p(b|x)$ from these rounds and uses them to estimate the correlator $I_{\mathrm{corr}} = |E_0 - E_1|$, thereby verifying that the devices exhibit nonclassical behaviour consistent with the energy constraint. 

An adversary, Eve, holds a quantum system $E$ that may be correlated with the preparation and measurement devices. Without loss of
generality, the joint state of Alice, Bob, and Eve before the channel application is a pure state
$\ket{\psi}_{PME}$.
After Alice applies $\Lambda_{x}$, the joint state becomes
\begin{equation}
  \rho^{x}_{SME}
  \;=\; \bigl(\Lambda_{x} \otimes \mathcal{I}_{ME}\bigr) \big[
        \ket{\psi}\!\bra{\psi}_{PME}\big]\,.
\end{equation}
Eve performs a POVM $\{Z^b_E\}_{b \in \{0,1\}}$ on her system,
attempting to guess Bob's outcome when $x=0$.  Given the protocol's constraints, the maximum probability that she guesses correctly is
\begin{equation}\label{eq:pguess}
  P_{\mathrm{guess}}(B|E, X=0)
  \;=\;
  \sup \;\sum_{b}\,
  \operatorname{Tr}\!\Big[
    \bigl(\Pi^b_{SM} \otimes Z^b_E\bigr)\,\rho^{0}_{SME}
  \Big],
\end{equation}
where the supremum is over all Hilbert spaces, pure states $\ket{\psi}_{PME}$, channels $\{\Lambda_x\}_x$, ground states $\ket{0}$, POVMs $\{\Pi^b_{SM}\}_b$, and Eve's POVMs $\{Z^b_E\}_b$, that are consistent with the energy constraint (Eq.~\eqref{eq:energy}) and the correlator constraint $I_{\mathrm{corr}} \geq I_{\mathrm{corr}}^{\exp}$.

The conditional min-entropy quantifying the extractable randomness
is
\begin{equation}\label{eq:Hmin}
  H_{\min}(B|E, X=0)
  \;=\; -\log_2 P_{\mathrm{guess}}(B|E, X=0)\,.
\end{equation}
This quantifies the number of bits, arbitrarily close to uniform and uncorrelated with Eve's system, that can be extracted from the raw outcomes by privacy amplification~\cite{renner2008security,tomamichel2011leftover}. Consequently, under the assumption of independent and identically distributed rounds, a certified lower bound on $H_{\min}$ (equivalently, an upper bound on $P_{\mathrm{guess}}$) constitutes a security guarantee which ensures that at least $H_{\min}$ bits of randomness can be extracted per round, regardless of the adversary's strategy.
In contrast, heuristic methods such as the seesaw optimisation employed in Refs.~\cite{d2025entanglement,roch2026role} identify particular feasible adversarial strategies, and therefore yield upper bounds on $H_{\min}$ (equivalently, lower bounds on $P_{\mathrm{guess}}$) and cannot rule out the existence of stronger, unexplored adversarial attacks. 


\ifarxiv\section{Certified lower bounds on $H_{\min}$ in the energy-constrained SDI scenario}\else\textit{Certified lower bounds on $H_{\min}$ in the energy-constrained SDI scenario.---}\fi Obtaining certified upper bounds on the guessing probability (Eq.~\eqref{eq:pguess}) is a challenging problem, not least because it involves an optimization over Hilbert spaces with an unbounded dimension. A standard way to overcome this is through outer-approximating the feasible set with a hierarchy of SDPs, namely, the NPA hierarchy~\cite{NPA1,navascues2008convergent}. This has been used to a great effect in the fully DI scenario (see, e.g.,~\cite{BRC,brown2024device}). The NPA hierarchy does not, however, apply to our scenario directly. Its standard formulation exploits the fact that the operators of spatially separated parties commute, whereas here Bob's POVM $\{\Pi^b_{SM}\}_b$ acts jointly on the message system $S$ and his local system $M$, while the energy assumption constrains $S$ alone. Source and measurement operators thus neither commute nor appear separately in the objective (see Eq.~\eqref{eq:pguess}). Our strategy is to resolve Bob's POVM in an orthonormal basis of $\mathcal{H}_S$, so that the source degrees of freedom appear explicitly and the resulting block operators, which act on $M$ alone, commute with them. We then build an NPA-style relaxation using these block operators.

We fix an arbitrary basis $\{\ket{s}\}_{s=0}^{d-1}$ for $\mathcal{H}_S$, with $d = \dim(\mathcal{H}_S)$, and choose $\ket{0}$ to be the ground state so that the energy constraint then reads $\bra{0}\rho^x_S\ket{0} \geq 1-\omega$. Recall that Bob applies a single binary POVM $\{\Pi^b_{SM}\}_{b\in\{0,1\}}$, which we can assume without loss of generality to be projective since any POVM on $SM$ can be realized as a projective measurement on an enlarged measurement system by taking its Naimark dilation, as we place no bound on $\dim(\mathcal{H}_{SM})$. We decompose $\Pi_{SM}^b$ across the $S$ and $M$ subsystems as
\begin{equation}\label{eq:POVM_decomp}
  \Pi^b_{SM}
  = \sum_{s_1,s_2=0}^{d-1}
    \ket{s_1}\!\bra{s_2}_S
    \otimes \Pi^b_M[s_1,s_2],
\end{equation}
where $\Pi^b_M[s_1,s_2] \in \mathcal{L}(\mathcal{H}_M)$ are operators on Bob's system $M$, whose dimension remains unspecified. Hermiticity of $\Pi^b_{SM}$ implies $\Pi^b_M[s_1,s_2] = (\Pi^b_M[s_2,s_1])^\dagger$, and completeness implies $\sum_b \Pi^b_M[s_1,s_2] = \delta_{s_1,s_2}\,\mathbb{I}_M$. Projectivity, in contrast, couples the blocks through an intermediate sum over the full basis,
\begin{equation}\label{eq:block_proj}
  \Pi^b_M[s_1,s_2]
  = \sum_{s_3=0}^{d-1}
    \Pi^b_M[s_1,s_3]\,\Pi^b_M[s_3,s_2],
\end{equation}
and therefore cannot be resolved into relations among a fixed set of blocks without reference to the value of $d$. Handling relations such as this in a dimension-independent way is the key technical step of our construction as explained next. We partition the decomposition of Eq.~\eqref{eq:POVM_decomp} into the blocks lying within the $\{\ket{0},\ket{1}\}$ subspace and a single auxiliary operator collecting all blocks with $s_1 \geq 2$ or $s_2 \geq 2$,
\begin{align}\label{eq:partition}
  \Pi^b_{SM}
  &= \sum_{s_1,s_2=0}^{1}
     \ket{s_1}\bra{s_2}_S \otimes \Pi^b_M[s_1,s_2]
  \nonumber\\
  &\quad+
  \underbrace{
    \sum_{\substack{(s_1,s_2):\\ s_1 \geq 2 \;\text{or}\; s_2 \geq 2}}
    \ket{s_1}\bra{s_2}_S \otimes \Pi^b_M[s_1,s_2]
  }_{=:\; F^b}.
\end{align}
The partition in Eq.~\eqref{eq:partition} preserves the two-dimensional subspace ($k=2$), which is used in all results presented below. We note that the construction generalizes to a preserved subspace of any dimension $k \geq 2$ (see Appendix~\ref{app:dimindep}). For a binary POVM, completeness fixes the $b=1$ operators, so we retain only $F := F^0$ together with the qubit-subspace operators $P := \ket{0}\!\bra{0}$, $Q := \ket{1}\!\bra{1}$ and $T := \ket{0}\!\bra{1}$, and the preserved blocks $A := \Pi^0_M[0,0]$, $B := \Pi^0_M[0,1]$ and $C := \Pi^0_M[1,1]$. Eve's binary projective measurement is likewise described by a single operator $Z := Z^0_E$.

We relax the tensor-product structure
between $M$ and $E$ to a commuting-operator model, in which operators
on different subsystems commute as
\begin{equation}\label{eq:all_comm}
  [O_S,\, O_{ME}] = 0, \quad
  [O_M,\, Z] = 0, \quad
  [F,\, Z] = 0,
\end{equation}
for all $O_S \in \{P, Q, T, T^\dagger\}$,
$O_{ME} \in \{A, B, B^\dagger, C, Z\}$, and
$O_M \in \{A, B, B^\dagger, C\}$.  Note that while the full POVM
$\Pi^b_{SM}$ does not commute with operators on $S$ (since it acts
jointly on $SM$), the individual blocks $\{A, B, B^\dagger, C\}$ do,
by virtue of the decomposition.  The auxiliary operator $F$ likewise
acts jointly on $SM$ and hence commutes with Eve's operator, but not, in general, with the qubit-subspace operators or with the blocks
$\{A, B, B^\dagger, C\}$. In addition, the auxiliary operator $F$ satisfies its own set of dimension-independent substitution rules.  By construction, $F$ has no support on the $\{0,1\} \times \{0,1\}$ sub-block of $S$.  Since each of the
operators $P$, $Q$, $T$, $T^\dagger$ corresponds to $\ket{i}\!\bra{j}$ for
some $i,j \in \{0,1\}$, sandwiching $F$ between any pair of them
gives
\begin{equation}\label{eq:sandwich}
  O_1\, F\, O_2 = 0,
  \qquad
  \forall\; O_1, O_2 \in \{P,\, Q,\, T,\, T^\dagger\}.
\end{equation}

In terms of the operators defined above, the POVM element
$\Pi^0_{SM}$ takes the form
\begin{equation}\label{eq:N_dimindep}
  N \;:=\; \Pi^0_{SM}
  \;=\; P\,A + T\,B + T^\dagger B^\dagger + Q\, C + F,
\end{equation}
with $\Pi^1_{SM} = \mathbb{I} - N$. For a given strategy, the guessing probability appearing in the objective of Eq.~\eqref{eq:pguess} reads $\operatorname{Tr}[(\mathbb{I} - N - Z + 2NZ)\,\rho^0]$, and $P_{\mathrm{guess}}$ is its supremum over all admissible strategies. We then introduce two moment matrices $\Gamma^0$ and $\Gamma^1$, one per prepared state, indexed by a common set of monomials in the operators $\mathcal{S}_1 = \{\mathbb{I}, P, T, T^\dagger, Q, A, B, B^\dagger, C, F, Z\}$, with entries $\Gamma^x_{u,v} = \operatorname{Tr}[\rho^x\, u^\dagger v] =: \langle u^\dagger v \rangle_x$. Collecting this objective together with the operator relations and all the constraints, as derived in detail in Appendix~\ref{app:dimindep}, we obtain the following noncommutative polynomial optimization problem:
\begin{widetext}
\begin{equation}\label{eq:NCPOP}
\boxed{
\begin{aligned}
  &\sup_{\mathcal{H},\;\rho^0,\rho^1 \in \mathcal{D}(\mathcal{H}),\;
  \{P,Q,T,F,A,B,C,Z\} \subset \mathcal{L}(\mathcal{H})}
  \quad
  \big\langle\, \mathbb{I} - N - Z + 2NZ \,\big\rangle_0\,,  
  \\[6pt]
  &\text{subject to:}\\[3pt]
  &
  \;\;\;\;\langle P \rangle_x \;\geq\; 1 - \omega,
  \quad x \in \{0,1\},
  \qquad
  2\langle N \rangle_0 - 2\langle N \rangle_1
  \;\geq\; I_{\mathrm{corr}}^{\exp}, \qquad \langle O \rangle_0 = \langle O \rangle_1
  \quad
  \forall\; O \in \{A,\, B,\, B^\dagger,\, C,\, Z\}^{\leq 2\ell},
  \\[3pt]
  & 
  \quad N^2 = N,
  \qquad
  Z^2 = Z = Z^\dagger,
  \qquad 0 \leq A,\, C,\, Z,\, N \leq \mathbb{I},
  \qquad
  P \succeq 0, \qquad Q \succeq 0, \qquad
  \mathbb{I} - P - Q \succeq 0,
  \\[3pt]
  &\quad P^2 = P,\quad Q^2 = Q,\quad
  T^2 = 0,\quad T\,T^\dagger = P,\quad
  T^\dagger T = Q,\quad
  P\,Q = 0,\quad
  P\,T = T,\quad T\,P = 0,\quad
  T\,Q = T,\quad Q\,T = 0,
  \\[3pt]
  &\quad O_1\, F\, O_2 = 0
  \qquad
  \forall\; O_1, O_2 \in \{P,\, Q,\, T,\, T^\dagger\}, \qquad [O_S,\, O_{ME}] = 0,
  \qquad
[O_M,\, Z] = 0,
  \qquad
  [F,\, Z] = 0.
\end{aligned}
}
\end{equation}
\end{widetext}
\noindent
Here $\langle O \rangle_x := \operatorname{Tr}[\rho^x\, O]$, and the supremum ranges over all Hilbert spaces $\mathcal{H}$, pairs of states $\rho^0, \rho^1 \in \mathcal{D}(\mathcal{H})$ and operator tuples in $\mathcal{L}(\mathcal{H})$ obeying the listed relations. $\ell$ is the NPA level and $\{A, B, B^\dagger, C, Z\}^{\leq 2\ell}$ denotes all monomials of degree at most $2\ell$ in these operators. The equalities $\langle O \rangle_0 = \langle O \rangle_1$ encode the fact that Alice's channel acts only on her share of the initial state, so the reduced state on $ME$ is independent of $x$ (see Appendix~\ref{app:dimindep} for details). We set $I^{\exp}_{\mathrm{corr}} = 4\sqrt{\omega(1-\omega)}$, the maximal correlator achievable without shared entanglement (Eq.~\eqref{eq:classical_bound}), i.e., the value Alice and Bob would record in an ideal implementation operating at energy $\omega$. Since the objective contains degree-3 monomials such as $PAZ$, a nontrivial bound requires $\ell \geq 2$, and hence all results reported below are obtained at $\ell = 2$.

Note that none of the relations in Eq.~\eqref{eq:NCPOP} reference the value of $d$. The resulting SDP is therefore a valid relaxation of Eq.~\eqref{eq:pguess} for \emph{every} finite source dimension $d \geq 2$ simultaneously and, since no bound is placed on $\dim(\mathcal{H}_M)$ or $\dim(\mathcal{H}_E)$ either, its optimum yields a certified lower bound on $H_{\min}$ (equivalently, a certified upper bound on $P_{\mathrm{guess}}$) involving no dimension assumption on any of the systems.

\begin{figure}[!t]
  \centering
  \includegraphics[width=0.8\linewidth]{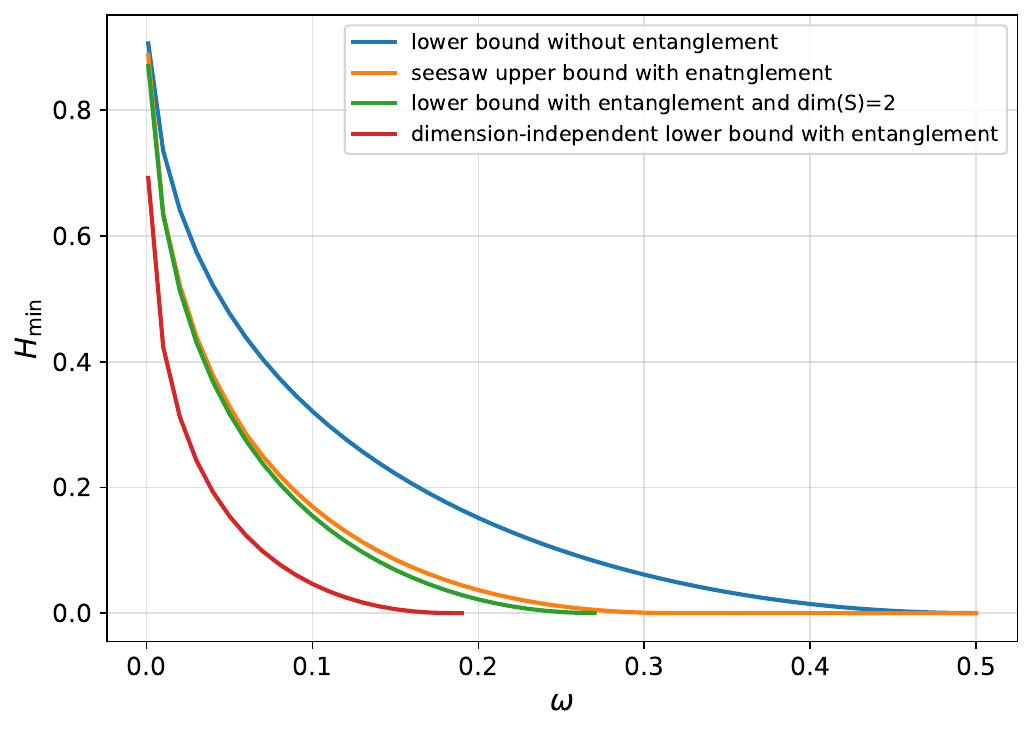}
  \caption{Certified min-entropy $H_{\min}$ as a function of the
  energy parameter $\omega$ without the dimension assumption (red curve). The blue curve shows the certified lower bound in the separable (no shared entanglement) scenario~\cite{van2019correlations}. The orange curve shows the upper bound obtained via the seesaw method in the entangled scenario with $\dim(\mathcal{H}_S) = \dim(\mathcal{H}_M) = 2$
\cite{d2025entanglement, roch2026role}.  The green curve shows our certified lower bound with $\dim(\mathcal{H}_S) = 2$ and $\dim(\mathcal{H}_M)$ unbounded.}
  \label{fig:dim_indep}
\end{figure}

Figure~\ref{fig:dim_indep} (red curve) shows the resulting certified min-entropy $H_{\min}$ as a function of the energy parameter $\omega$.  The dimension-independent bound is strictly positive for $\omega \lesssim 0.18$ and vanishes beyond this threshold, establishing the central claim of this work: certified randomness generation is feasible in the energy-constrained SDI scenario against a fully quantum adversary, without any assumption on the dimensions of the underlying quantum systems. As expected, the bound lies below the certified rate of the scenario without shared entanglement~\cite{van2019correlations} (blue curve), since an entangled adversary is strictly more powerful~\cite{d2025entanglement, roch2026role}. Our dimension-independent certified rates also lie below the seesaw upper bound~\cite{d2025entanglement, roch2026role} (orange curve). For comparison, Fig.~\ref{fig:dim_indep} also shows a certified lower bound obtained using our approach under the additional qubit-source assumption (green curve), while the dimensions of $M$ and $E$ remain unrestricted. This additional assumption yields a higher certified rate that lies closer to the seesaw upper bound. The corresponding fixed-source-dimension relaxation is detailed in Appendix~\ref{app:fixeddim}. 

Although our numerical benchmarks use the ideal correlator as the acceptance threshold, exact saturation is not required. By continuity, the certified min-entropy remains positive under a sufficiently small reduction of this threshold, demonstrating nonzero tolerance to experimental imperfections. For instance, at $\omega = 0.05$, the certified min-entropy remains positive for observed correlator values down to $58\%$ of the ideal value, demonstrating substantial tolerance to imperfections in the observed correlator. Further details on the robustness of the dimension-independent bounds are provided in Appendix~\ref{app_subsec:noise}.

Finally, we note that the relaxation of Eq.~\eqref{eq:NCPOP} is not tied to the guessing probability. Replacing the objective by the correlator itself, and dropping Eve's operators and the correlator constraint, which then play no role, turns the same construction into a certified upper bound on the largest correlator attainable at a given energy when arbitrary shared entanglement is permitted. This yields an outer approximation of the set of achievable correlations, complementing the inner approximations obtained from the explicit entangled strategies of Refs.~\cite{d2025entanglement,roch2026role}. We report these results in Appendix~\ref{app:icorr}.

\ifarxiv\section{Conclusion}\else\textit{Conclusion.---}\fi We have constructed SDP relaxations based on the NPA hierarchy for the guessing probability in the energy-constrained
prepare-and-measure scenario, allowing for arbitrary shared entanglement between the preparation and measurement devices.  Our relaxations provide the first certified lower bounds on the extractable randomness in this setting, resolving the open question highlighted in Refs.~\cite{van2019correlations,d2025entanglement,roch2026role, bhavsar2026higher} of whether any randomness can be rigorously certified against a fully quantum adversary.
Our dimension-independent lower bound vanishes beyond $\omega\approx0.18$. In comparison, the seesaw upper bounds of Refs.~\cite{d2025entanglement,roch2026role} remain positive up to $\omega\approx0.32$, while randomness can be certified throughout $\omega<1/2$ when the devices share only classical correlations~\cite{van2019correlations}. The true rate lies between our lower bound and the seesaw upper bounds. Closing this gap, through tighter relaxations or stronger explicit attacks, remains an open problem. Furthermore, our bounds hold for arbitrarily large finite dimensions, with no upper bound imposed. Extending them to infinite-dimensional systems, a natural setting for a photonic implementation, remains an open question.

Several other directions remain open for future investigation. Symmetry reduction techniques, for instance, exploiting the $\mathrm{U}(d{-}1)$ symmetry acting on the excited subspace of $\mathcal{H}_S$, could significantly reduce the SDP size and make higher NPA hierarchy levels computationally accessible. Moreover, because our relaxation is a noncommutative polynomial optimization over fixed operator sets, with the two prepared states linked by moment equalities, the Brown--Fawzi--Fawzi method applies essentially as a drop-in replacement of the objective, enabling certified bounds on the conditional von Neumann and Petz--R\'enyi entropies in this energy-constrained SDI scenario~\cite{brown2024device,hahn2024bounds}. Also, extending our single-round analysis to a finite-round security proof would
bring the results closer to practical SDI QRNG implementations. 

More broadly, our framework can directly incorporate experimentally determined bounds on individual photon numbers, since these enter the moment relaxation as linear constraints. The vacuum and one-photon populations are already represented by the operators $P$ and $Q$, respectively, while higher photon-number components can be included by enlarging the explicitly retained source subspace. This would extend the photon-number-constrained approach of Ref.~\cite{roch2025prepare}, which assumes classically correlated devices and classical side information, to the present setting with pre-shared entanglement between the devices and arbitrary quantum side information held by Eve.

\ifarxiv\section*{Code availability}\else\textit{Code availability.---}\fi
The code used to obtain the numerical results and generate the figures is publicly available at \url{https://github.com/Shashankranu/Hmin_energy-const_SDI}

\ifarxiv\section*{Note added}\else\textit{Note added.---}\fi
While preparing this manuscript, we became aware of an independent and concurrent work~\cite{li2026chip} that also obtains dimension-independent certified min-entropy bounds for an energy-constrained prepare-and-measure device in the presence of entanglement between the preparation and measurement devices, and applies them to an SDI randomness amplification protocol.

\ifarxiv\section*{Acknowledgments}\else\textit{Acknowledgments.---}\fi This project has received funding from the European Union’s Horizon Europe research and innovation programme under the project ``Quantum Secure Networks Partnership" (QSNP, grant agreement No 101114043), from ChistEra-2023/05/Y/ST2/00005 under the project Modern Device Independent Cryptography (MoDIC) and from the PEPR integrated project DIQKD ANR-22-PETQ-0009 as part of Plan France 2030.

\bibliography{ref}

\onecolumngrid
\appendix

\section{Details of the dimension-independent relaxation}
\label{app:dimindep}
Recall that we decompose $\Pi_{SM}^b$ across the $S$ and $M$ subsystems as
\begin{equation}\label{eq:app_POVM_decomp}
  \Pi^b_{SM}
  = \sum_{s_1,s_2=0}^{d-1}
    \ket{s_1}\!\bra{s_2}_S
    \otimes \Pi^b_M[s_1,s_2],
\end{equation}
where $d = \dim(\mathcal{H}_S)$ and
$\Pi^b_M[s_1,s_2] \in \mathcal{L}(\mathcal{H}_M)$ are operators on the measurement subsystem, whose dimension remains unspecified. The structural constraints on the POVM translate into algebraic relations among the blocks of Eq.~\eqref{eq:app_POVM_decomp}. Hermiticity $\Pi^b_{SM} = (\Pi^b_{SM})^\dagger$ implies
\begin{equation}\label{eq:block_herm}
  \Pi^b_M[s_1,s_2]
  = \big(\Pi^b_M[s_2,s_1]\big)^\dagger.
\end{equation}
Projectivity $\Pi^b_{SM} = (\Pi^b_{SM})^2$ gives
\begin{equation}\label{eq:block_proj}
  \Pi^b_M[s_1,s_2]
  = \sum_{s_3=0}^{d-1}
    \Pi^b_M[s_1,s_3]\,\Pi^b_M[s_3,s_2].
\end{equation}
Completeness $\sum_b \Pi^b_{SM} = \mathbb{I}_{SM}$ yields
\begin{equation}\label{eq:block_complete}
  \sum_b \Pi^b_M[s_1,s_2]
  = \delta_{s_1,s_2}\,\mathbb{I}_M.
\end{equation}
Note that, unlike the Hermiticity and completeness relations, the projectivity relation couples the blocks through a sum over the intermediate index $s_3$ that runs over the full range $\{0,\ldots,d-1\}$.

For a fixed strategy, we denote by $p_{\mathrm{guess}}$ the objective function of Eq.~\eqref{eq:pguess}, whose supremum over all admissible strategies is $P_{\mathrm{guess}}$, and by $I_{\mathrm{corr}} = E_0 - E_1$ the correlator difference entering the constraint. Both are linear in the block operators as shown
\begin{equation}\label{eq:obj_blocks}
p_{\mathrm{guess}} = 
  \sum_b \operatorname{Tr}\!\big[
    \rho^0_{SME}\,
    (\Pi^b_{SM} \otimes Z^b_E)\big]
  = \sum_b \sum_{s_1,s_2}
    \operatorname{Tr}\!\Big[\rho^0_{SME}\,
    \big(\ket{s_1}\!\bra{s_2}_S
    \otimes \Pi^b_M[s_1,s_2]
    \otimes Z^b_E\big)\Big],
\end{equation}

and

\begin{equation}\label{eq:corr_blocks}
I_{\mathrm{corr}} = 
  \sum_{x,b} (-1)^{x\oplus b}\,
\operatorname{Tr}\!\big[\rho^x_{SM}\,\Pi^b_{SM}\big]= \sum_{x,b} (-1)^{x\oplus b} \sum_{s_1,s_2}
    \operatorname{Tr}\!\Big[\rho^x_{SM}\,
    \big(\ket{s_1}\!\bra{s_2}_S
    \otimes \Pi^b_M[s_1,s_2]\big)\Big].
\end{equation}
Every term appearing in these expressions corresponds to an entry of a moment matrix in the relaxation.

Next, to arrive at our dimension-independent relaxation, we partition the decomposition of $\Pi_{SM}^b$  as shown

\begin{equation}
    \Pi^b_{SM}
  = \sum_{s_1,s_2=0}^{1}
     \ket{s_1}\bra{s_2}_S \otimes \Pi^b_M[s_1,s_2]
  \quad+
  \underbrace{
    \sum_{\substack{(s_1,s_2):\\ s_1 \geq 2 \;\text{or}\; s_2 \geq 2}}
    \ket{s_1}\bra{s_2}_S \otimes \Pi^b_M[s_1,s_2]
  }_{=:\; F^b}. \label{eq:app_partition}
\end{equation}

Hermiticity of $\Pi^b_{SM}$ implies $F^b = (F^b)^\dagger$ for the auxiliary operator of Eq.~\eqref{eq:app_partition}, and $A$ and $C$ are Hermitian while $B$ is in general non-Hermitian. For general $d$, the projector onto the orthogonal complement of $\mathrm{span}\{\ket{0},\ket{1}\}$, $R := \mathbb{I}_S - P - Q$, is a nonzero positive operator, giving the constraint $\mathbb{I}_S - P - Q \succeq 0$ of Eq.~\eqref{eq:NCPOP}. The orthonormality of $\ket{0}$ and $\ket{1}$ yields the substitution rules
\begin{equation}\label{eq:S_subs_gen}
\begin{aligned}
  &P^2 = P, \quad Q^2 = Q, \quad
   T^2 = (T^\dagger)^2 = 0,\\[2pt]
  &T\,T^\dagger = P, \quad T^\dagger T = Q,\\[2pt]
  &P\,T = T, \quad T\,P = 0, \quad
   P\,T^\dagger = 0, \quad T^\dagger P = T^\dagger,\\[2pt]
  &P\,Q = Q\,P = 0,\\[2pt]
  &T\,Q = T, \quad Q\,T = 0, \quad
   T^\dagger Q = 0, \quad Q T^\dagger = T^\dagger,
\end{aligned}
\end{equation}
all of which hold for every $d \geq 2$. 

The auxiliary operator $F$ satisfies its own set of dimension-independent substitution rules.  By construction, $F$ has no support on the $\{0,1\} \times \{0,1\}$ sub-block of $S$.  Since each of the
operators $P$, $Q$, $T$, $T^\dagger$ corresponds to $\ket{i}\!\bra{j}$ for
some $i,j \in \{0,1\}$, sandwiching $F$ between any pair of them
gives
\begin{equation}\label{eq:sandwich}
  O_1\, F\, O_2 = 0,
  \qquad
  \forall\; O_1, O_2 \in \{P,\, Q,\, T,\, T^\dagger\}.
\end{equation}
These sixteen degree-3 relations hold for every $d \geq 2$ and make no reference to the value of $d$.

Eve's POVM is binary and projective, with the single independent operator $Z := Z^0_E$ satisfying $Z^2 = Z$, $Z = Z^\dagger$, and $Z^1_E = \mathbb{I} - Z$.  Relaxing the tensor-product structure between $M$ and $E$ to a commuting-operator model gives
\begin{equation}\label{eq:all_comm}
  [O_S,\, O_{ME}] = 0, \quad
  [O_M,\, Z] = 0, \quad
  [F,\, Z] = 0,
\end{equation}
for all $O_S \in \{P, Q, T, T^\dagger\}$, $O_{ME} \in \{A, B, B^\dagger, C, Z\}$, and $O_M \in \{A, B, B^\dagger, C\}$. While the full POVM $\Pi^b_{SM}$ does not commute with operators on $S$, the individual blocks $\{A, B, B^\dagger, C\}$ do, by virtue of the decomposition. The auxiliary operator $F$ acts jointly on $SM$ and hence commutes with Eve's operator, but not, in general, with the qubit-subspace operators or with the blocks.

The guessing probability under this commuting-operator model is $p_{\mathrm{guess}} = \sum_b \operatorname{Tr}[N_b\, Z_b\, \rho^0]$ with $N_0 = N$, $N_1 = \mathbb{I} - N$, $Z_0 = Z$, $Z_1 = \mathbb{I} - Z$, and expands as
\begin{align}\label{eq:pguess_expand}
  p_{\mathrm{guess}}
  &= \operatorname{Tr}\!\big[
     (1 - N - Z + 2NZ)\,\rho^0\big].
\end{align}

We next impose the constraints that link the two moment matrices.
Alice's channel $\Lambda_x$ acts only on the system $P$, so tracing
over $S$ yields a reduced state on $ME$ that is independent of $x$,
\begin{equation}\label{eq:nosig}
  \operatorname{Tr}_S[\rho^0_{SME}]
  = \operatorname{Tr}_P[\ket{\psi}\!\bra{\psi}_{PME}]
  = \sigma_{ME}
  = \operatorname{Tr}_S[\rho^1_{SME}].
\end{equation}
This no-signalling property implies that for any operator $O$
acting only on the $ME$ subsystem,
\begin{equation}\label{eq:ME_eq}
  \langle O \rangle_0 = \langle O \rangle_1.
\end{equation}
We impose Eq.~\eqref{eq:ME_eq} for every monomial $O$ of the
operators $\{A, B, B^\dagger, C, Z\}$ of degree at most $2\ell$,
where $\ell$ is the NPA level, and all such monomials are expressible
as entries of the level-$\ell$ moment matrices.  These equalities
are linear constraints linking corresponding entries of $\Gamma^0$
and $\Gamma^1$. 

The energy constraint can be written as
\begin{equation}\label{eq:energy_moment}
  \operatorname{Tr}[\rho^x  P] \;\geq\; 1 - \omega
  \quad \forall\; x \in \{0,1\},
\end{equation}
which is a linear inequality on the entry $\Gamma^x_{1,P}$.  For
the correlator, since the POVM is binary, we have
$E_x = 2\langle N \rangle_x - 1$, so that
$E_0 - E_1 = 2\langle N \rangle_0 - 2\langle N \rangle_1$.  Each $\langle N \rangle_x$ is a
sum of the terms
$\langle PA \rangle_x$, $\langle TB \rangle_x$,
$\langle T^\dagger B^\dagger \rangle_x$,
$\langle Q C \rangle_x$, and $\langle F \rangle_x$, which are entries in the moment matrix $\Gamma^x$, and the constraint
$E_0 - E_1 \geq I_{\mathrm{corr}}^{\exp}$ becomes
\begin{equation}\label{eq:corr_constraint}
  2\langle N \rangle_0 - 2\langle N \rangle_1
  \;\geq\; I_{\mathrm{corr}}^{\exp},
\end{equation}
a single linear inequality linking $\Gamma^0$ and $\Gamma^1$.  Here
$I_{\mathrm{corr}}^{\exp} = 4\sqrt{\omega(1-\omega)}$ is the bound on the correlator when Alice and Bob do not share entanglement.

We impose the projectivity of the full POVM element, $N^2 = N$, at the level of moments. Note that the operator equality $N^2 = N$ is equivalent to the pair of operator inequalities $\pm(N^2 - N) \succeq 0$. Imposed through localizing matrices at level $\ell$, this pair forces every entry of the corresponding localizing matrices to vanish, i.e., it is equivalent to the family of linear moment equalities
\begin{equation}\label{eq:proj_moment}
  \big\langle\, O_1\,(N^2 - N)\,O_2 \,\big\rangle_x \;=\; 0, \qquad x \in \{0,1\},
\end{equation} 
over monomials $O_1, O_2$ of degree at most $2\ell$. In our implementation we enforce Eq.~\eqref{eq:proj_moment} for $O_1, O_2 \in \{\mathbb{I}, P, Q, T, T^\dagger\}$ to improve computational performance. Removing constraints only enlarges the feasible set, so any such truncation is conservative, and the optimum remains a valid certified upper bound on $P_{\mathrm{guess}}$. Finally, we impose the  remaining positivity constraints
\begin{equation}\label{eq:positivity}
\begin{aligned}
  &0 \leq A \leq \mathbb{I}, \quad
  0 \leq C \leq \mathbb{I}, \quad
  0 \leq Z \leq \mathbb{I}, \\[2pt]
  &0 \leq N \leq \mathbb{I}, \quad
  P \succeq 0, \quad Q \succeq 0, \quad
  \mathbb{I} - P - Q \succeq 0,
\end{aligned}
\end{equation}
through localizing matrix constraints. All relaxations were generated using the Python package \textsc{ncpol2sdpa}~\cite{wittek2015algorithm} and solved with \textsc{mosek}~\cite{mosek}.

We note that the decomposition of $\Pi_{SM}^b$ shown in Eq.~\eqref{eq:app_partition} can be extended to a preserved subspace $\mathrm{span}\{\ket{0}, \ldots, \ket{k-1}\}$ of any dimension $k \geq 2$. This requires introducing projectors $Q_j = \ket{j}\!\bra{j}$ and transition operators $T_{ij} = \ket{i}\!\bra{j}$ for $i,j < k$, retaining the blocks $\Pi^0_M[i,j]$ with $i,j < k$, and collecting all remaining blocks into a single auxiliary operator $F$ satisfying the sandwich rules $O_1 F O_2 = 0$ for all $O_1, O_2$ in the preserved-subspace operator set, together with the positivity constraint $\mathbb{I}_S - \sum_{j<k} Q_j \succeq 0$. Taking $k = 2$ recovers the construction described in detail above.

\section{Details of relaxation under fixed source dimension assumption}
\label{app:fixeddim}
The dimension-independent relaxation above assumes nothing about any Hilbert-space dimension, which is what makes its output a security guarantee in the strongest sense. It is nevertheless critical to explore how much this generality costs, and to compare directly with the existing literature, where the upper bounds on the min-entropy are obtained assuming the dimension of both systems $S$ and $M$ to be qubits~\cite{d2025entanglement,roch2026role}.  Hence, we fix the dimension of the message system $S$, while leaving everything else unconstrained. We emphasize that, even under this assumption, no dimension bound is placed on system $M$ and Eve's system $E$. This distinguishes our setting from Refs.~\cite{d2025entanglement,roch2026role}, which fix the dimension of the joint system $SM$ while obtaining upper bounds on the min-entropy. By leaving these systems unbounded, we grant the adversary strictly more freedom, resulting in a more conservative security analysis.

Setting $d = 2$ produces two immediate simplifications of the
dimension-independent relaxation shown in Eq.~\eqref{eq:NCPOP}. First, there are no basis
indices $s \geq 2$, so the auxiliary operator $F$ vanishes, and the sandwich rules of Eq.~\eqref{eq:sandwich}
become vacuous.  Second, the basis $\{\ket{0},\ket{1}\}$ is now
complete, so $Q = \ket{1}\!\bra{1} = \mathbb{I}_S - P$ and the
operator $Q$ can be eliminated. Every occurrence of $Q$ in
Eq.~\eqref{eq:S_subs_gen} is replaced by $\mathbb{I}_S - P$, giving
in particular
\begin{equation}\label{eq:TdT_fixed}
  T^\dagger T = \mathbb{I}_S - P.
\end{equation}
Also, the complement projector $R = \mathbb{I}_S - P - Q$
vanishes.  The POVM element of Eq.~\eqref{eq:N_dimindep}
correspondingly reduces to
\begin{align}\label{eq:N_assembled}
  N &= P\,A + T\,B + T^\dagger B^\dagger
     + (\mathbb{I}_S - P)\,C
  \nonumber\\
  &= P(A - C) + T\,B + T^\dagger B^\dagger + C,
\end{align}
with $\Pi^1_{SM} = \mathbb{I} - N$.

The main simplification of fixing the source dimension is that projectivity becomes easier to impose. With $F = 0$, the intermediate sum in Eq.~\eqref{eq:block_proj} is restricted to $s_3 \in \{0,1\}$ and closes within the known blocks. The projectivity condition can therefore be resolved at the block level, yielding the degree-2 substitution rules
\begin{equation}\label{eq:proj_rules}
\begin{aligned}
  A^2 &= A - B\,B^\dagger,
    &\quad& (0,0)\text{ block},\\[2pt]
  C^2 &= C - B^\dagger B,
    &\quad& (1,1)\text{ block},\\[2pt]
  A\,B &= B - B\,C,
    &\quad& (0,1)\text{ block},\\[2pt]
  B^\dagger A &= B^\dagger - C\,B^\dagger,
    &\quad& (1,0)\text{ block}.
\end{aligned}
\end{equation}

The moment matrices $\Gamma^0$ and $\Gamma^1$ are now built from the
reduced operator set
\begin{equation}\label{eq:S1_fixed}
  \mathcal{S}_1 = \{
  \mathbb{I},\; P,\; T,\; T^\dagger,\;
  A,\; B,\; B^\dagger,\; C,\; Z\},
\end{equation}
consisting of 9~elements, which produces a $9\times 9$ moment matrix
at NPA level~1.  As before, the objective contains degree-3
monomials, so we work at NPA level~2.  The remaining constraints
carry over unchanged from Eq.~\eqref{eq:NCPOP}. 

Figure~\ref{fig:dim_indep} (green curve) in the main text shows the lower bound on $H_{\min}$ as a function of $\omega$ for the qubit source at NPA level~2.  The min-entropy is strictly positive for $\omega \lesssim 0.27$ and vanishes beyond this threshold.  In contrast, the seesaw upper bounds of~\cite{d2025entanglement} report a positive min-entropy up to $\omega \approx 0.32$.  This gap is attributable to two factors. First, our bounds are certified lower bounds on $H_{\min}$ that may not be tight, whereas the seesaw provides only an upper bound. Part of the gap may therefore be relaxation slack rather than the existence of genuinely stronger attacks. Second, we leave the measurement system $M$ unbounded, whereas~\cite{d2025entanglement} fixes both $S$ and $M$ to be two-dimensional. The larger adversarial space offered by an unbounded $M$, in principle, permits stronger attacks and hence may lead to lower certified rates.

The same construction generalises to higher fixed source dimensions.
For $\dim(\mathcal{H}_S) = 3$, the POVM decomposes into $d^2 = 9$
blocks per outcome, and one introduces transition operators
$\ket{i}\!\bra{j}$ for $i,j \in \{0,1,2\}$, yielding a larger
operator set and a correspondingly richer collection of substitution
rules and ME-equality constraints.  Table~\ref{tab:dimensions}
summarises the key parameters of the relaxation for the qubit and
qutrit source dimension assumptions.

\begin{figure}[!hbtp]
  \centering
  \includegraphics[width=0.45\linewidth]{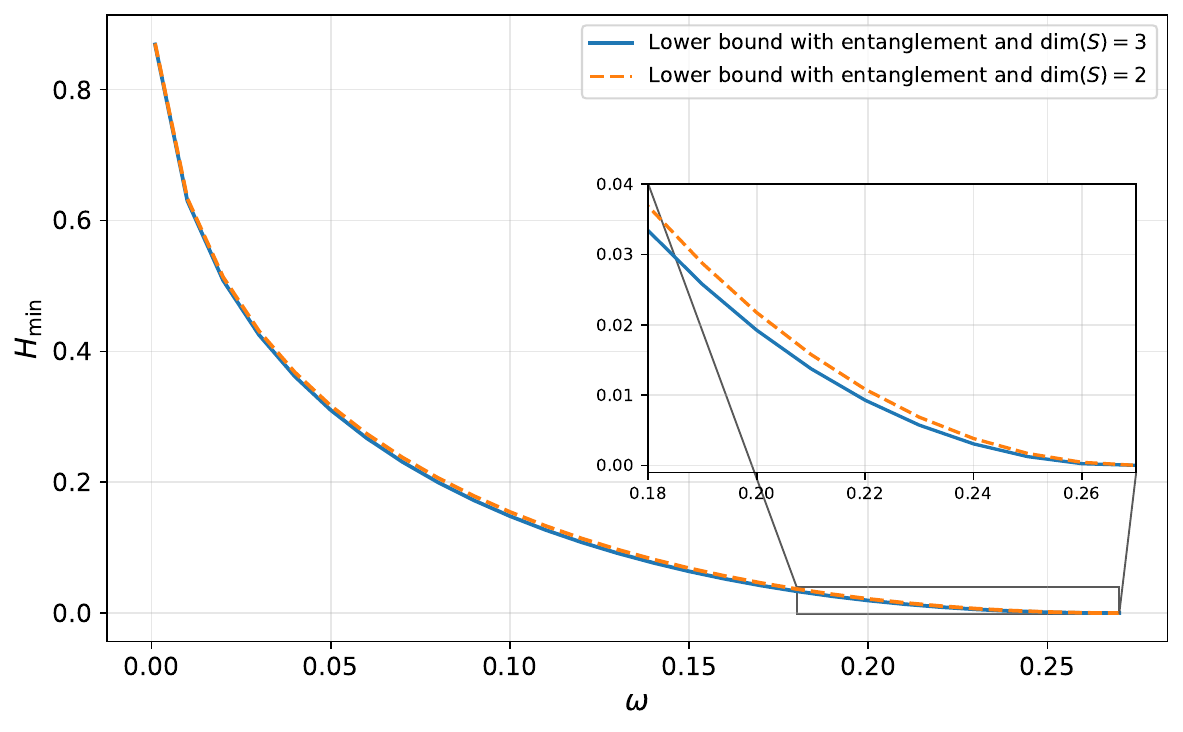}
  \caption{Certified lower bounds on $H_{\min}$ as a function of the energy parameter $\omega$ for source dimensions $\dim(\mathcal{H}_S) = 2$ (orange) and $\dim(\mathcal{H}_S) = 3$ (blue).}
  \label{fig:qutrit}
\end{figure}

We also compare the certified min-entropy as a function of the energy parameter $\omega$ for the qubit and qutrit source dimension assumptions in Fig.~\ref{fig:qutrit}. Both curves are obtained from our fixed-dimension relaxation at NPA level~2 with $\dim(\mathcal{H}_M)$ unbounded. The two curves nearly coincide across the full range of $\omega$, suggesting that increasing the source dimension beyond a qubit provides the adversary with little additional advantage.  This observation complements the dimension-independent analysis, providing numerical evidence that the optimal guessing probability itself depends weakly on the source dimension. It also suggests that the gap between the fixed-dimension and dimension-independent bounds in Fig.~\ref{fig:dim_indep} may not come mainly from stronger higher-dimensional attacks. Instead, part of the gap may be due to the weaker moment-level enforcement of the projectivity constraint in the dimension-independent relaxation or due to the fact that both bounds are computed at $\ell = 2$.

\begin{table}[t]
\centering
\caption{Moment matrix size for qubit and qutrit source dimension assumptions.}
\label{tab:dimensions}
\begin{tabular}{@{}ccc@{}}
\toprule
$d$ & NPA level & $\dim(\Gamma^x)$ \\
\midrule
2 & 1 & $9\times 9$   \\
2 & 2 & $40\times 40$ \\
3 & 1 & $19\times 19$ \\
3 & 2 & $180\times 180$ \\
\bottomrule
\end{tabular}
\end{table}

\section{Robustness to imperfections in the observed statistics}\label{app_subsec:noise}
The results reported in the main text take the observed correlator to attain the maximal value achievable without shared entanglement, $I^{\exp}_{\mathrm{corr}} = 4\sqrt{\omega(1-\omega)}$, corresponding to an ideal implementation at energy $\omega$. Any experimental imperfection will reduce the observed value, and it is therefore important that randomness remains certifiable below it. This is straightforward to assess with our relaxation, since $I^{\exp}_{\mathrm{corr}}$ enters Eq.~\eqref{eq:NCPOP} only through a single linear constraint. We therefore fix $\omega$ and solve the dimension-independent SDP at $\ell = 2$ over a range of $I^{\exp}_{\mathrm{corr}}$ values up to the ideal value. 

\begin{figure}[!hbtp]
  \centering
  \begin{subfigure}[t]{0.45\linewidth}
    \centering
    \includegraphics[width=\linewidth]{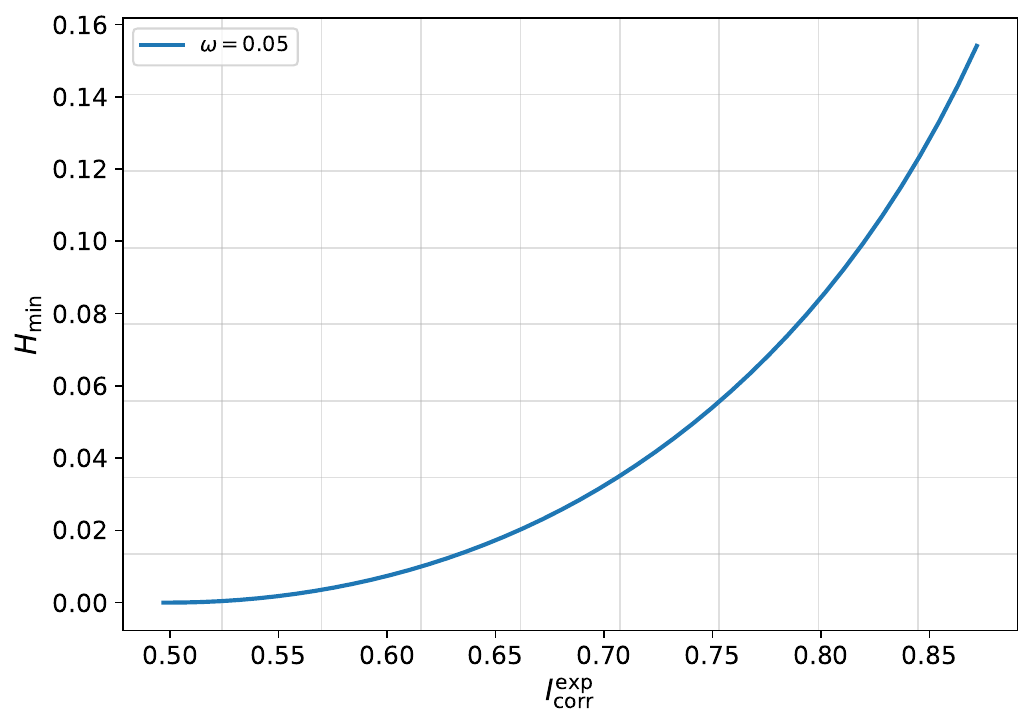}
    \caption{}
    \label{fig:noise005}
  \end{subfigure}
  \hfill
  \begin{subfigure}[t]{0.45\linewidth}
    \centering
    \includegraphics[width=\linewidth]{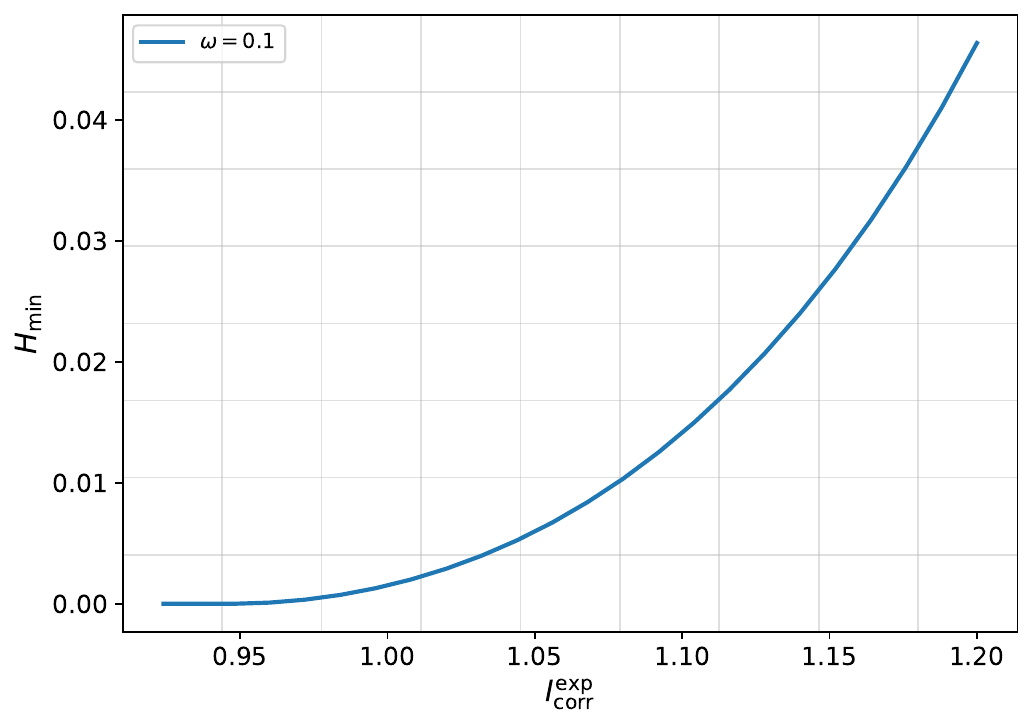}
    \caption{}
    \label{fig:noise01}
  \end{subfigure}
  \caption{Certified min-entropy $H_{\min}$ from the dimension-independent relaxation as a function of the observed correlator $I^{\exp}_{\mathrm{corr}}$ at (a) $\omega = 0.05$ and (b) $\omega = 0.1$. The ideal correlator values are approximately 0.87 and 1.2, respectively, attained at the right-hand end of each curve.}
  \label{fig:noise}
\end{figure}

Figure~\ref{fig:noise} shows the certified min-entropy as a function of $I^{\exp}_{\mathrm{corr}}$ at $\omega = 0.05$ and $\omega = 0.1$. In both cases $H_{\min}$ decreases monotonically as the observed correlator is reduced and vanishes at a value strictly below the ideal one. For $\omega = 0.05$, the bound remains positive down to $I^{\exp}_{\mathrm{corr}} \approx 0.50$, compared with the ideal value of $\approx 0.87$. Similarly, for $\omega=0.1$, the ideal correlator value is $1.2$, and positive min-entropy is certified above approximately $0.948$, corresponding to about $79\%$ of the ideal value. Operating at low energy is therefore doubly advantageous, yielding both a higher certified rate and greater robustness to imperfections in the observed statistics.

\section{Certified upper bounds on the maximal correlator}
\label{app:icorr}

Our relaxations can be modified to bound the maximal correlator achievable under the energy constraint, complementing the lower bounds obtained via explicit entangled strategies in Refs.~\cite{d2025entanglement,roch2026role}. Concretely, we consider
\begin{equation}\label{eq:Icorr_opt}
  I^{\max}_{\mathrm{corr}}(\omega)
  \;=\; \sup\; 2\langle N \rangle_0 - 2\langle N \rangle_1,
\end{equation}
subject to the normalisation, energy, ME-equality, projectivity and positivity constraints of Eq.~\eqref{eq:NCPOP}. Eve's operators play no role in this optimisation and are dropped. Since the feasible set of the relaxation contains all physical realisations, the SDP optimum is a certified upper bound on $I^{\max}_{\mathrm{corr}}(\omega)$, valid for arbitrary source dimension in the dimension-independent variant, and under the corresponding dimension assumption in the fixed-$d$ variant.

\begin{figure}[!hbtp]
  \centering
  \includegraphics[width=0.4\linewidth]{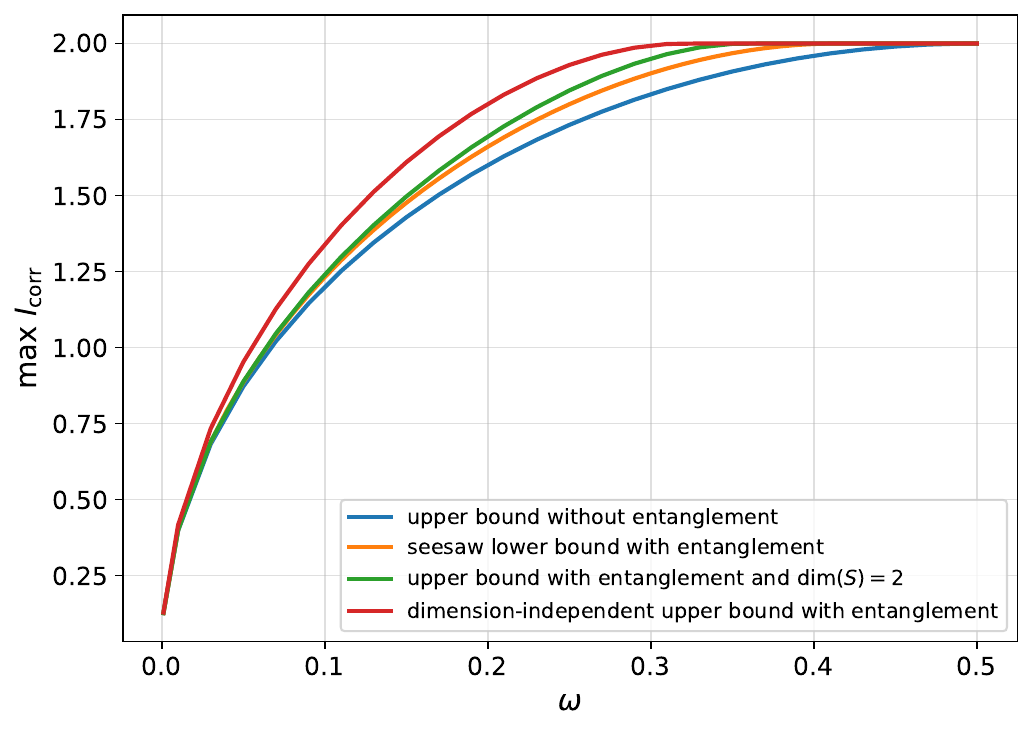}
  \caption{Certified upper bounds on the maximal correlator $I_{\mathrm{corr}}$ as a function of the energy parameter $\omega$, obtained from the dimension-independent relaxation (red curve) and the fixed-dimension relaxations with $d=2$ (green curve), compared with the maximal correlator without shared entanglement (blue curve) $I^{\mathrm{rand}}_{\mathrm{corr}}(\omega) = 4\sqrt{\omega(1-\omega)}$~\cite{van2017semi} and the explicit-strategy lower bounds (orange curve) of Ref.~\cite{d2025entanglement}.}
  \label{fig:icorr}
\end{figure}

The explicit strategies of Ref.~\cite{d2025entanglement} already show that shared entanglement can surpass the no-entanglement limit $I^{\mathrm{rand}}_{\mathrm{corr}}(\omega)=4\sqrt{\omega(1-\omega)}$. Our bounds complement this result by placing a certified upper limit on the maximal entanglement-assisted value. As shown in Fig.~\ref{fig:icorr}, the explicit-strategy lower bounds (orange curve) and dimension-independent upper bounds (red curve) lie close together, confining the true maximum to a narrow interval throughout the considered range of $\omega$. Our fixed-dimension upper bound (green curve) is lower than the dimension-independent one (red curve), mirroring the behaviour of the min-entropy bounds in the main text and consistent with the weaker moment-level enforcement of projectivity in the dimension-independent relaxation.

\end{document}